\documentclass[a4paper]{article}
\usepackage{graphicx}
\title{100 ps time-of-flight resolution \\of Dielectric Resistive Plate Chamber}
\author{A.~Akindinov$^1$, \and V.~Golovin$^2$, \and A.~Martemiyanov$^1$, \and V.~Petrov$^2$, \and V.~Plotnikov$^1$, \and A.~Smirnitsky$^1$, \and K.~Voloshin$^{1,}$\footnote{Corresponding author. E-mail: Kirill.Voloshin@itep.ru}}
\date{}
\begin{document}
\thispagestyle{empty}
\maketitle

\begin{center}
$^1${\it Institute for Theoretical and Experimental Physics (ITEP), \\B.~Cheremushkinskaya 25, Moscow, 117218, Russia.}\\
$^2${\it Center of Perspective Technologies and Apparatus (CPTA), \\Preobrazhenskaya pl. 6/8, Moscow, 107076, Russia.}
\end{center}

\begin{abstract}
Time of flight of a minimum ionizing particle along a fixed base 
has been measured with a 100~ps accuracy by means of a Dielectric 
Resistive Plate Chamber (DRPC) with $4 \times 0.3$~mm gas gaps. DRPC 
timing characteristics have been studied with different applied 
voltages, discriminating thresholds and beam intensities. 
It may be stated that the time-of-flight resolution of 
gaseous detectors developed within the ALICE experiment has reached 
the level of the best known scintillation counters. 
\end{abstract}

During the last several years a revolutionary progress in breakdown 
suppression inside gaseous time-of-flight detectors was achieved by 
introducing, in different ways, a resistivity inside the gas gap 
\cite{patent,itep4598,CERN}. Despite of this fact, there was a poor idea 
about timing 
properties, which may be principally manifested by these detectors. 

A value of 100~ps seems to be a natural limit in this sence. Basing 
on the ALICE physical conditions, 100~ps time-of-flight resolution is 
sufficient for $\pi$/K/p separation in a real momenta range. 
Untill recently, such fine resolution could be provided only by 
modern scintillation counters and Pestov spark counters \cite{Pestov}. 
As an example, the timing system based on scintillators and 
photomultipliers, proposed for the STAR project at RHIC, provides 
the time resolution of about 90~ps \cite{STAR}.

\begin{figure}
\begin{center}
\includegraphics*[width=.7\linewidth]{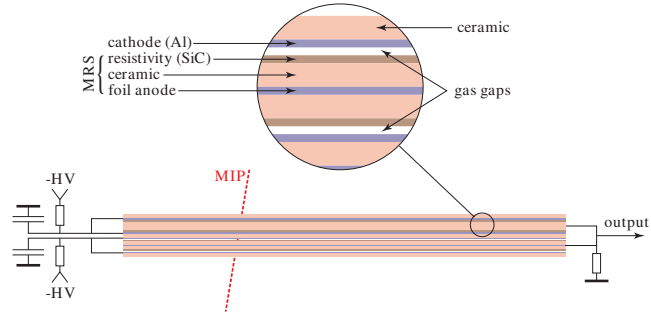}
\end{center}
\caption{DRPC construction.}
\label{DRPC}
\end{figure}

The detector, described in the given paper, 
is schematically presented in Fig.~\ref{DRPC}. Dielectric Resistive Plate
Chamber (DRPC) consists of several ceramic plates (0.5~mm of ordinary
unpolished ceramics) which form four gas gaps, each 0.3~mm wide. In accordance 
with the expectations, decreasing the gap width has led to the rise 
of the time
resolution. The number of gaps (two in previous version \cite{itep4598}) 
was doubled to keep the MIP registration efficiency close to 100\%. 

The chamber consists of two types of electrodes. Ceramic cathodes are
metallized with aluminum. Dielectric-resistive electrodes are also
made of ceramics metallized with aluminum on the one side and covered,
through evaporation, with semi-conducting SiC on the other side. The plates
are assembled in pairs so that the metal layer, common for two gaps,
is positioned inside, and the semi-conducting layers are turned towards the
gaps. The idea of electrical connection between the semi-conductor and
the metal is described in Ref.~\cite{itep4598}. The detector has a square
working surface of $2 \times 2$~cm$^2$. 

Methods employed in the measurements of TOF resolution and registration
efficiency at ITEP and CERN accelerators was, in general, similar to
that decribed in Ref.~\cite{itep4598}. The same 
front-end electronics and gas mixture consisting of 
85\%C$_2$H$_2$F$_4$ + 5\%isobutane + 10\%SF$_6$ were used. 
The start part of the setup, based on scintillation counters, was 
modified, so that the information from several detectors could be analized
simultaneously. 

The counting rate, or efficiently registered particle flux over the
working surface, is an important parameter of DRPC. 
During the measurements the rate was fixed at the level of
1~kHz$ \cdot $s$^{-1} \cdot $cm$^{-2}$, which is higher than that predicted
under the ALICE conditions
(100--200~Hz$\cdot$s$^{-1}\cdot$cm$^{-2}$). Special measurements of
the way the time resolution dependends on the rate were performed as well.

\begin{figure}[h]
\begin{center}
\begin{tabular}{cc}
(a) & (b) \\
\includegraphics*[width=.45\linewidth]{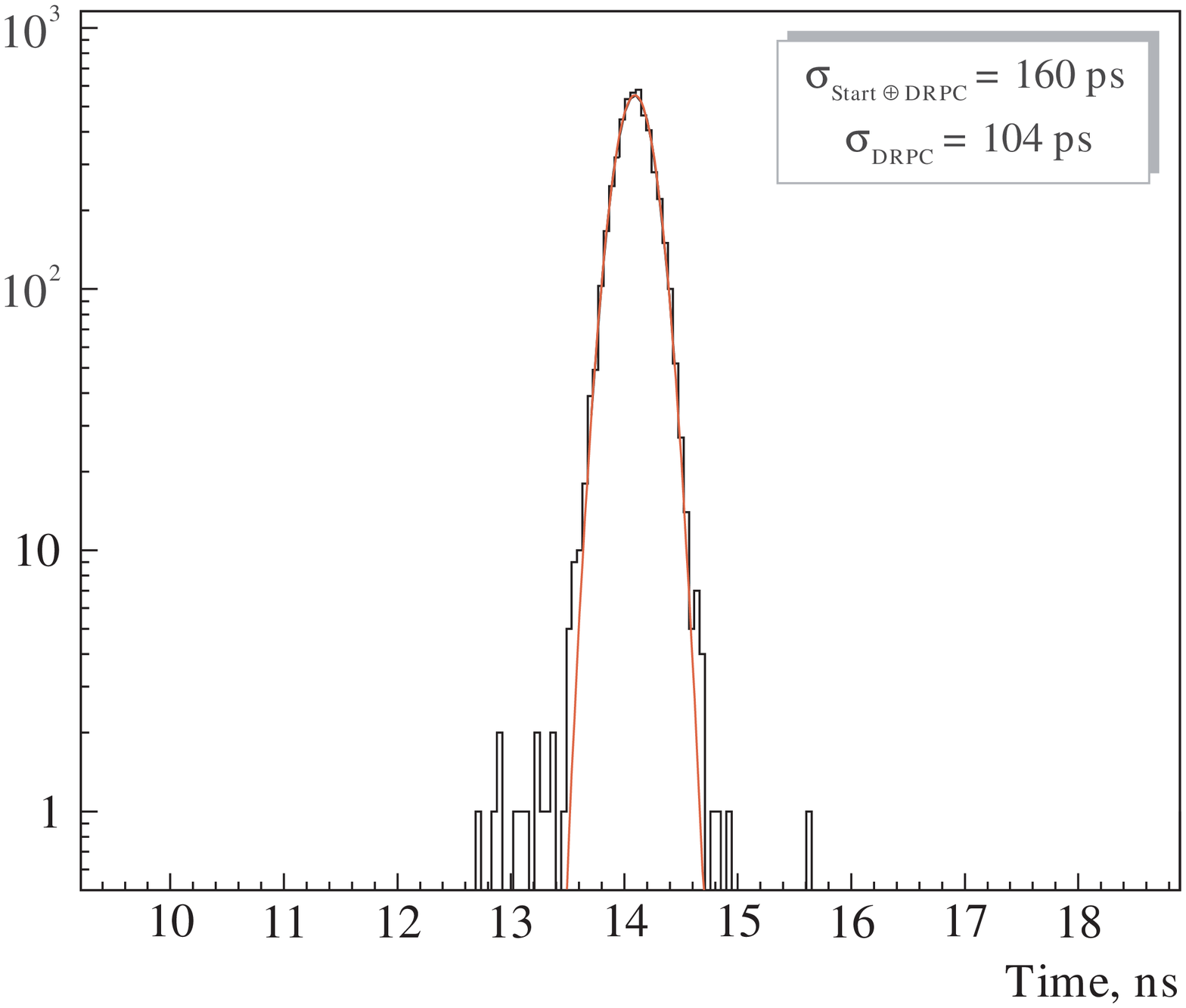} &
\includegraphics*[width=.45\linewidth]{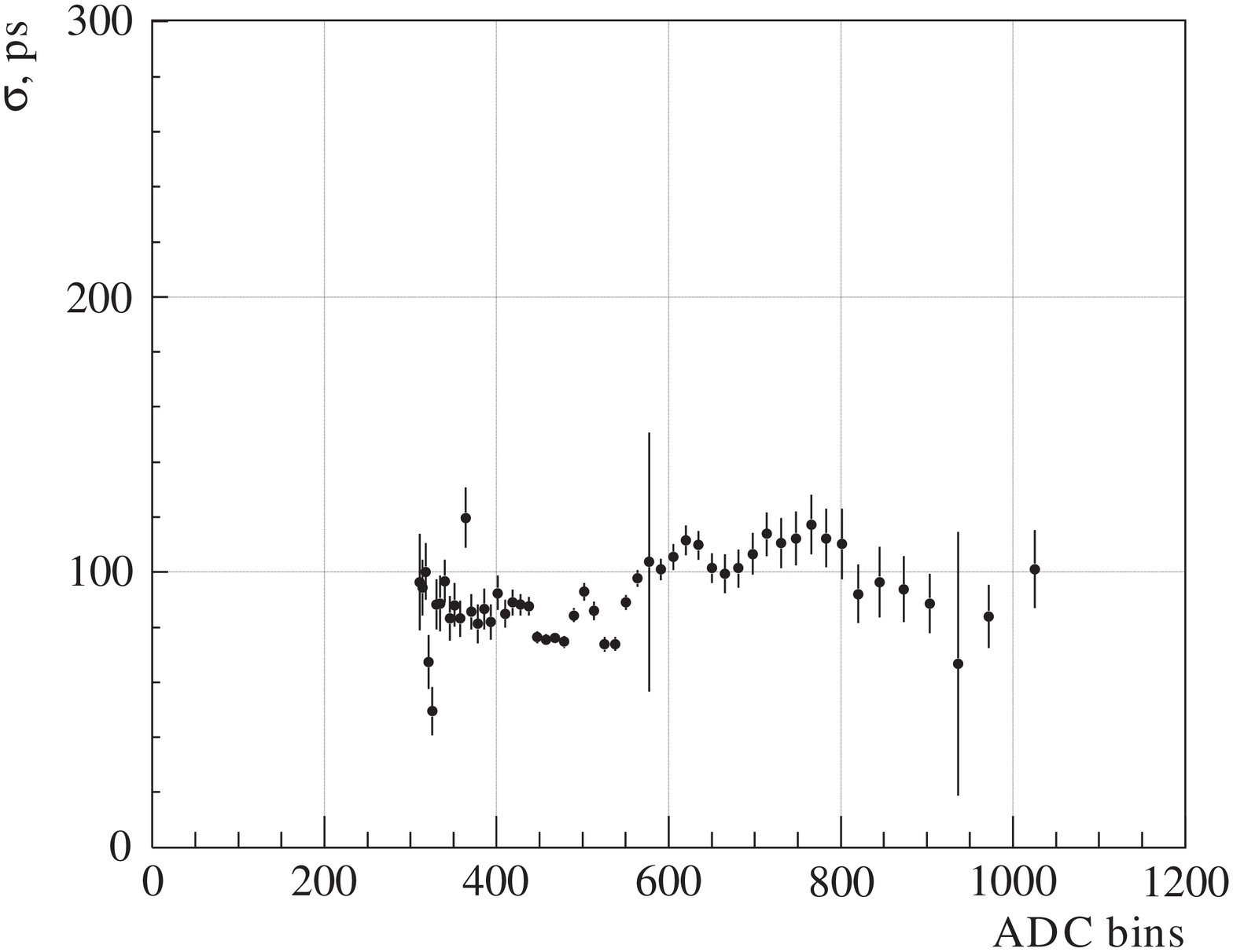} \\
\end{tabular}
\end{center}
\caption{DRPC time resolution: (a) total,
(b) for different ranges in amplitudes.}
\label{tof}
\end{figure}

A typical time-of-flight distribution, summarised over the whole
range of amplitudes, with the total registration efficincy close to
95\%, is shown in Fig.~\ref{tof}a. One can see that the standard
deviation is really on the level of 100~ps, and the distribution has
unsignificant {\it tails}. Fig.~\ref{tof}b represents the
same data in more detail: the time resolution is shown for different
amplitudes. The dependence is very slight, the
resolution stays close to 100~ps in the whole range of amplitudes, 
which explains the absence of {\it tails} in Fig.~\ref{tof}a. 

\begin{figure}[h]
\begin{tabular}{cc}
(a) & (b) \\
\includegraphics*[width=.45\linewidth]{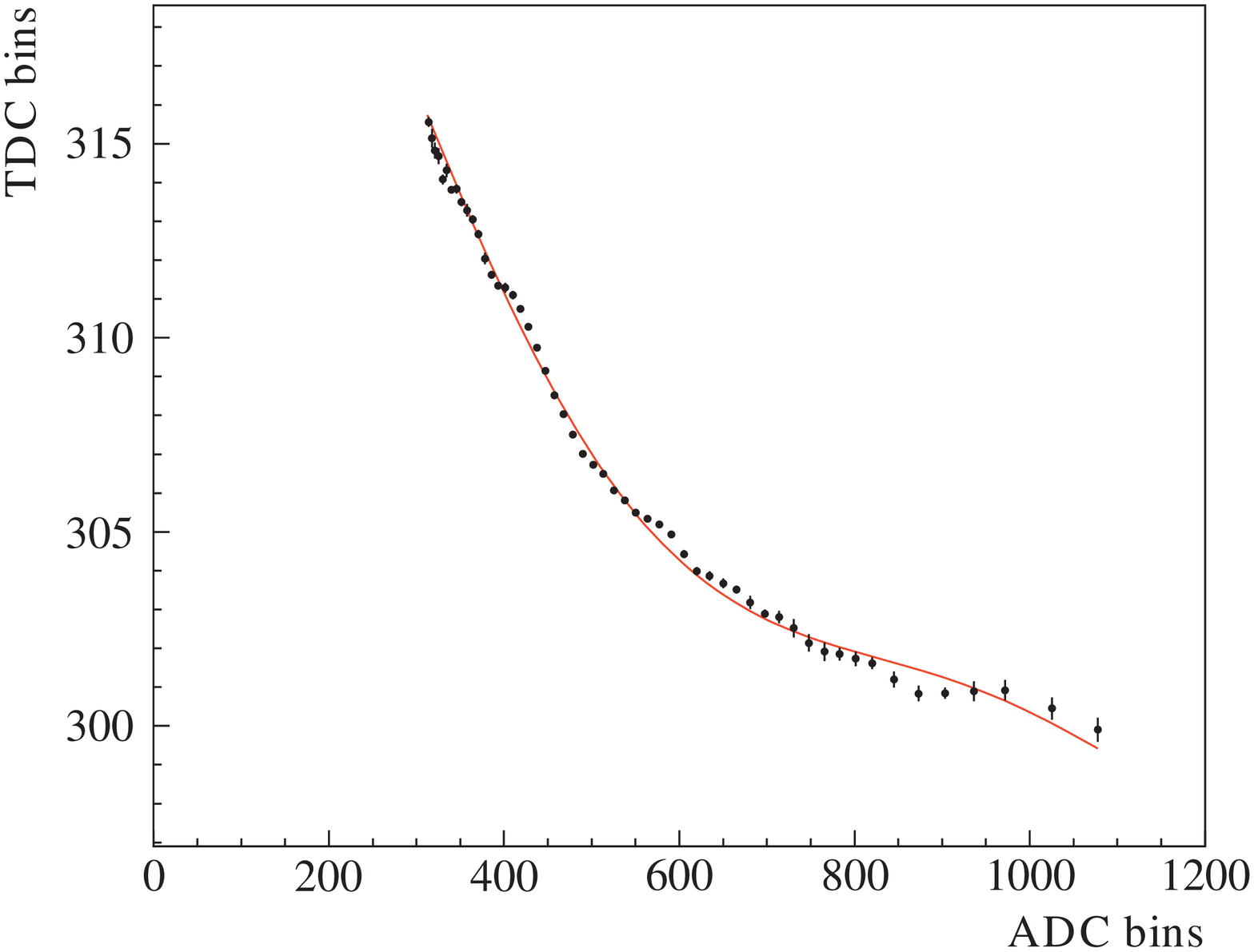} &
\includegraphics*[width=.45\linewidth]{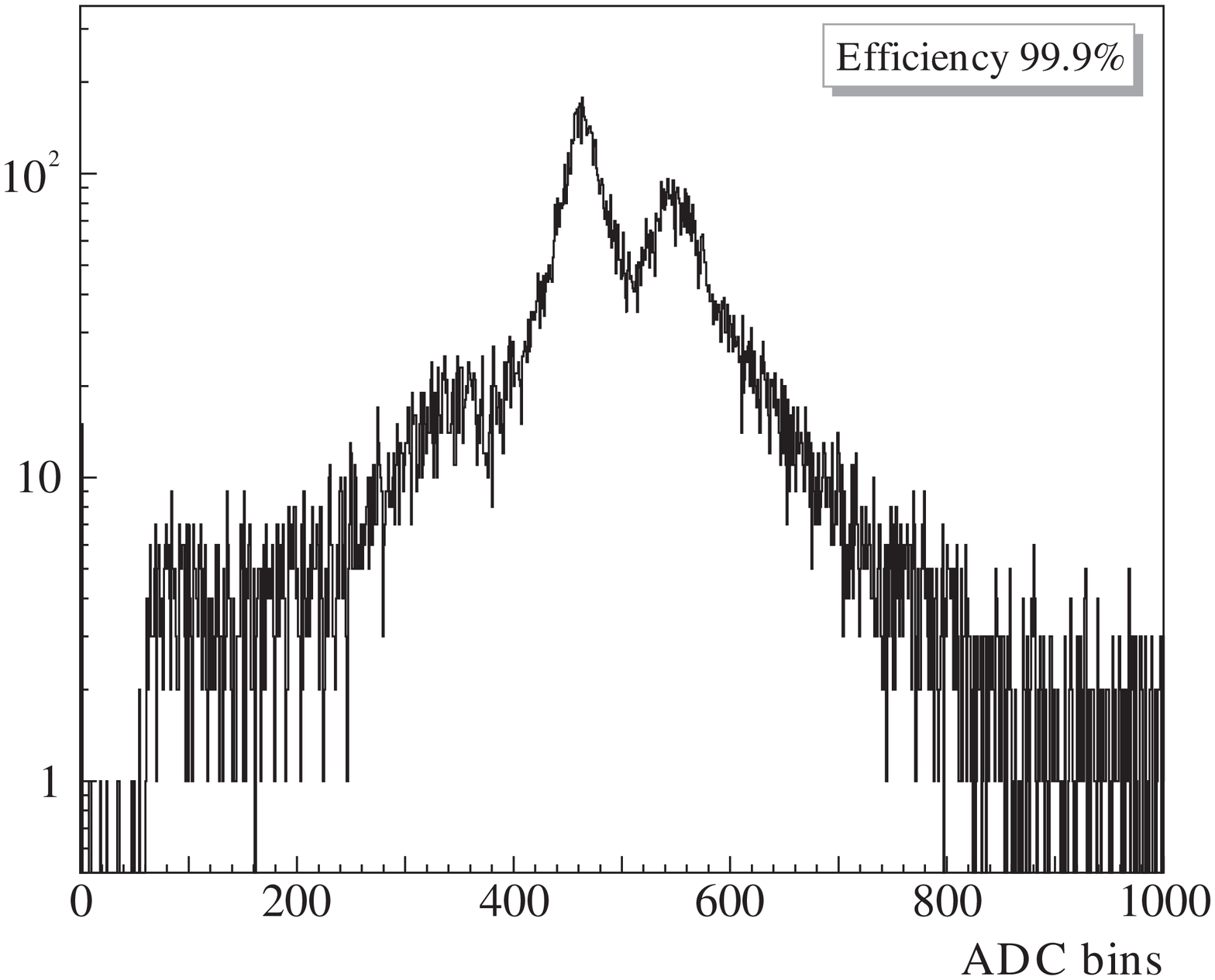} \\
\end{tabular}
\caption{(a) Polynomial slewing correction and (b) overall DRPC amplitude 
spectrum.}
\label{amp}
\end{figure}

Actually, the timing resolution is calculated after slewing correction, 
which takes into account the fact that signals with larger amplitudes
are triggered by the constant-threshold discriminator at earlier times. 
Such a correction influences the timing distribution in a
strong way. It is normally performed with a polynomial function in a way
shown in Fig.~\ref{amp}a.  

\begin{figure}[h]
\begin{center}
\includegraphics*[width=.7\linewidth]{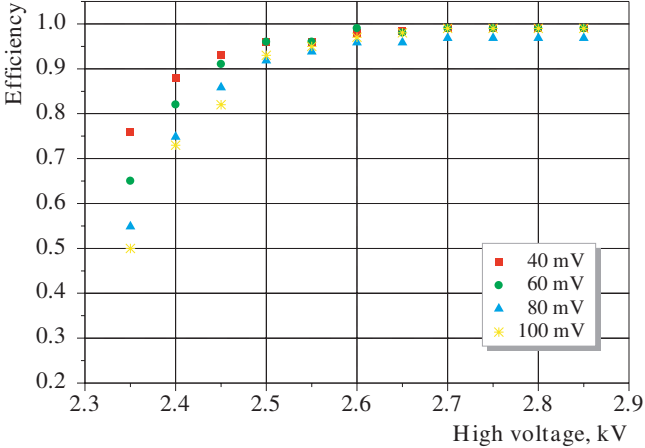}
\end{center}
\caption{DRPC efficiency for different high voltages and discriminator
  thresholds.} 
\label{etof}
\end{figure}

The amplutude spectrum obtained from the amplifier output is shown in
Fig.~\ref{amp}b. More specifically, it shows the charge
integrated by a charge-sencitive ADC. Although the front-end
electronics is not linear in the whole range of amplitudes, it may be
seen that the amplitude spectrum has a peak, staying far
from the pedestal bounder on a slightly changing background. Amplitude
magnitudes correspond to the gas amplification of 10$^7$ and allow to
obtain excellent registration efficiency at different high voltages
and electronics thresholds. The dependence of efficiency on the
high voltage is shown in Fig.~\ref{etof} for different electronics
thresholds. Even at the threshold of 100~mV there is a clear plateau, in which 
the efficiency stays close to 100\%.

\begin{figure}[h]
\begin{tabular}{cc}
(a) & (b) \\
\includegraphics*[width=.45\linewidth]{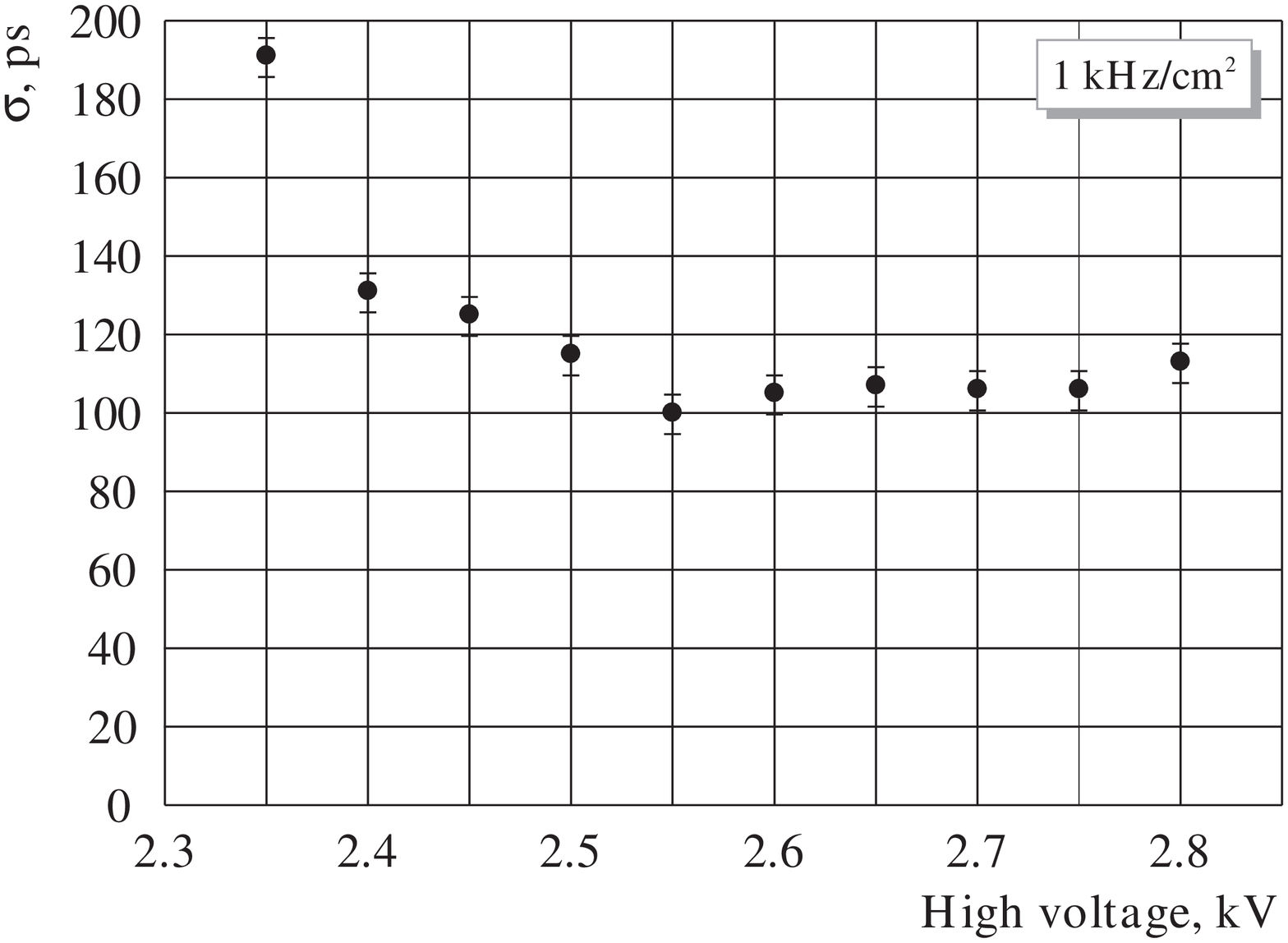} &
\includegraphics*[width=.45\linewidth]{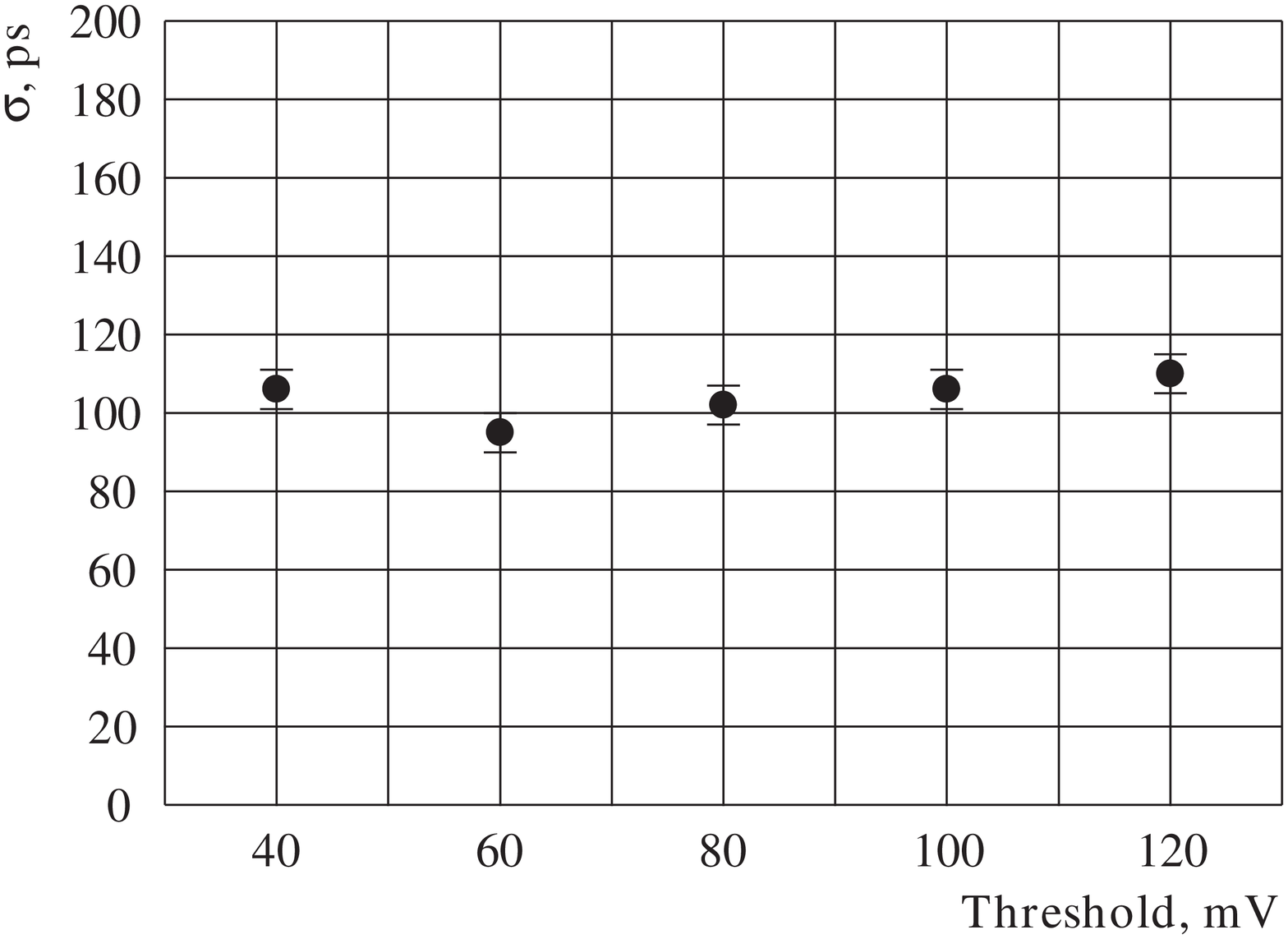} \\
\end{tabular}
\caption{DRPC time-of-flight resolution (a) at different high voltages
and (b) discriminating thresholds.} 
\label{hvth}
\end{figure}

The fact that the TOF resolution does not depend on the
high voltage and the discriminating threshold, is illustrated in 
Fig.~\ref{hvth}.

\begin{figure}[h]
\begin{center}
\includegraphics*[width=.7\linewidth]{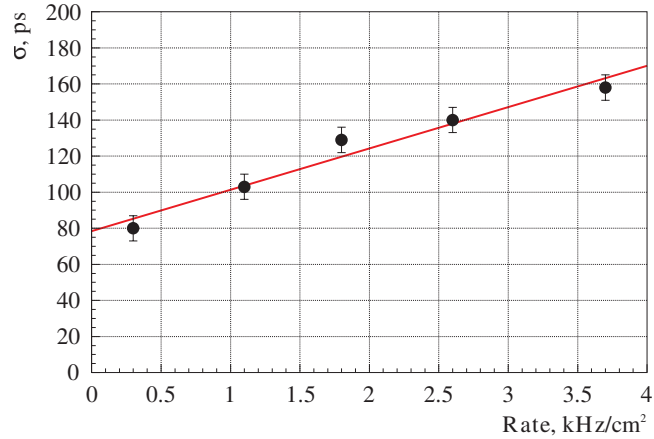}
\end{center}
\caption{DRPC TOF resolution at different counting rates.} 
\label{rate}
\end{figure}

\begin{figure}
\begin{center}
\includegraphics[width=.7\linewidth]{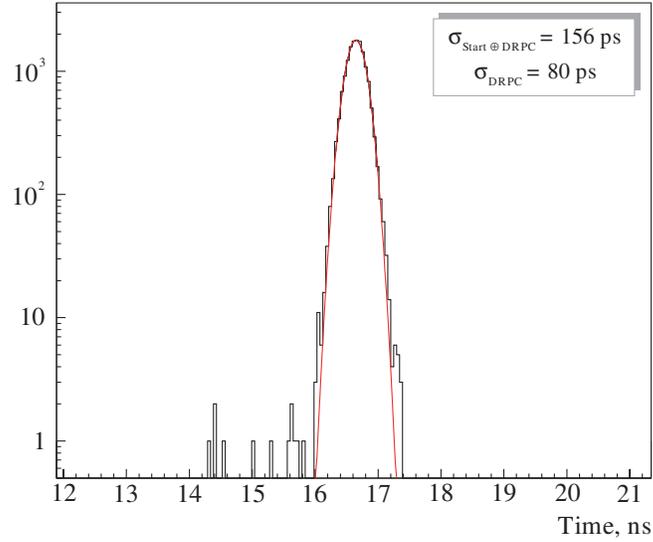}
\end{center}
\caption{DRPC time resolution at a low counting rate.}
\label{fresol80}
\end{figure}

All the results described above were obtained at a fixed counting rate. A
special experiment was performed to study the rate influence on the
detector properties. Fig.~\ref{rate} shows the dependence of the TOF
resolution on the particle flux at 40~mV electronics threshold. The
resolution increases with the rate up to approximately 150~ps. But under
the ALICE conditions (the very beginning of the scale) it may be
expected to be as low as 80~ps. Fig.~\ref{fresol80} shows the low rate
resolution in detail. The distribution is still very clear with the
{\it tails} admixture being less than 10$^{-3}$.

The study of DRPC timing properties was performed with a support from
RFFI grant \#99--02--18377.

\end{document}